\documentclass[preprints,article,accept,moreauthors,pdftex]{Definitions/mdpi} 

\usepackage{bm}
\usepackage{bbm}
\usepackage{listings}
\usepackage{subfigure}
\usepackage{amsmath}
\usepackage{amssymb}
\usepackage{amsthm}
\usepackage{booktabs}
\usepackage{hyperref}
\usepackage{erewhon} 
\usepackage[sf,scale=1]{libertine}
\usepackage{fourier}
\usepackage[T1]{fontenc}
\usepackage{algorithm}
\usepackage{algorithmic}

\hypersetup{colorlinks,citecolor=teal}

\firstpage{1} 
\pubvolume{xx}
\issuenum{1}
\articlenumber{1}
\pubyear{2022}
\copyrightyear{202}
\history{}

\newcommand*\diff{\mathop{}\!\kern0pt\mathrm{d}}

\Title{How to choose my stochastic volatility parameters? A review}
\Author{Fabien Le Floc'h}
\AuthorNames{Fabien Le Floc'h}
\address{}
\corres{Correspondence: fabien@2ipi.com}
\abstract{Based on the existing literature, this article presents the different ways of choosing the parameters of stochastic volatility models in general, in the context of pricing financial derivative contracts. This includes the use of stochastic volatility inside stochastic local volatility models.}

\begin{document}



\section{The major stochastic (local) volatility models}
The calibration practices presented in this paper are applicable to the most common stochastic volatility model. Let us recap the main models:
\begin{itemize}
	\item In the Heston stochastic local volatility model, an asset $F$ is described by the following system of stochastic differential equations \cite{heston1993closed}:
\begin{subequations}	\begin{align}\label{eqn:heston_forward_process}
		\frac{\diff F}{F} &=  \sqrt{v}  L(F,t)\diff W_F\,,\\
		\diff v &= \kappa (\theta - v)\diff t + \sigma \sqrt{v} \diff W_v		
	\end{align} \end{subequations}
	with $\diff W_F \diff W_v = \rho \diff t$. For an equity of spot $S$, and maturity $T$, $F$ represents the forward to maturity $F(t)=S(t) e^{(r-q)(T-t)}$ with $r$ the relevant interest rate, and $q$ the dividend yield. The function $L(S,t)$ is an additional local volatility component (also called leverage function). In the pure stochastic volatility model, $L = 1$. Some issues of the Heston model are the not so realistic variance distribution when the Feller condition is not verified (which is common when applied with $L=1$ in practice), the relatively poor fit due to the limited number of parameters, and the lack of flexibility in the forward variance dynamic. Its main advantage is to have an affine characteristic function, which allows for an efficient calculation of vanilla option prices. The pure stochastic volatility fit can be greatly increased through the use of piecewise-constant vol-of-variance $\sigma$  and correlation $\rho$ in time, which is usually accompanied by the choice $v(0)=1=\theta$ and some exogenously fixed speed of mean reversion $\kappa$ \citep{hagan2018implied}.
	\item The Double-Heston model of \cite{christoffersen2009shape} is more flexible with regards to the fit and the dynamic. The asset follows
\begin{subequations}	\begin{align}\label{eqn:double_heston_sde}
		\frac{\diff F}{F} &=  \sqrt{v_1}  L(F,t)\diff W_{F_1} + \sqrt{v_2} L(F,t)\diff W_{F_2}\,,\\
		\diff v_1 &= \kappa_1 (\theta_1 - v_1)\diff t + \sigma_1 \sqrt{v_1} \diff W_{v_1}\,,\\
		\diff v_2 &= \kappa_2 (\theta_2 - v_2)\diff t + \sigma_2 \sqrt{v_2} \diff W_{v_2}\,,
	\end{align} \end{subequations}
	where $W_{F_1}$,$W_{v_1}$ are two Brownian motions correlated with correlation $\rho_1$, $W_{F_2}$,$W_{v_2}$ are two Brownian motions correlated with correlation $\rho_2$ and each pair $(W_{v_1}, W_{v_2})$, $(W_{F_1}, W_{F_2})$, $(W_{v_1}, W_{F_2})$, $(W_{F_1}, W_{v_2})$ is uncorrelated. Because of the added dimension, the particle method of \citet{guyon2011smile} may be necessary to obtain the local volatility component from the standard Dupire local volatility and price exotics. When used as pure stochastic volatility model, in this classic formulation, obtaining stable parameters over time is challenging.
	\item In the Bates model, a jump component is added, and we have
\begin{subequations}	\begin{align}
		\frac{\diff F}{F} &= -\lambda \bar{k} \diff t + \sqrt{v} \diff W_F + \diff Z\,,\\
		\diff v &= \kappa(\theta - v)\diff t + \sigma \sqrt{v} \diff W_v\,,
\end{align} \end{subequations}
	where $W_F$ and $W_v$ are Brownian motions correlated with correlation $\rho$, $\lambda$ is the annual frequency of jumps, $k$ is the random percentage jump conditional on a jump occurring, and $Z$ is a Poisson process with intensity $\lambda$ and log-normal distribution of jumps sizes $k$ such that $\ln(1+k) \sim N(\ln(1+\bar{k})-\frac{1}{2}\delta^2, \delta^2)$.
	\item  The Schobel-Zhu model is another type of affine model where the volatility $v$ is a mean-reverting Ornstein-Uhlenbeck process \cite{schobel1999stochastic}:
\begin{subequations}	\begin{align}\label{eqn:schobel_zhu_forward_process}
		\frac{\diff F}{F} &=  v  L(F,t)\diff W_F\,,\\
		\diff v &= \kappa (\theta - v)\diff t + \sigma \diff W_v\,,
	\end{align} \end{subequations}
	with $\diff W_F \diff W_v = \rho \diff t$. In this model, the volatility process can become negative. While it is not a mathematical issue, as this means that the correlation sign is simply flipped, the flipping of the correlation sign is more of a concern from an economic perspective. Fortunately, the probability of it being negative is very low in practice, has a negligible effect  on the price of derivatives \citep{zhu2009applications}. Still, on paths where the stochastic volatility is negative, it is stays negative for an unrealistic period of time \citep{healy2025applied}. 
	\item The exponential Ornstein-Uhlenbeck model of \citet{scott1987option}
\begin{subequations}	\begin{align}
		\frac{\diff F}{F} &=  v  L(F,t)  \diff W_F\,,\\
		\diff \ln v &= \kappa (\theta - \ln v)\diff t + \sigma \diff W_v\,,
	\end{align} \end{subequations}
	The model is not affine but European options may be priced via a one dimensional Monte-Carlo simulation \citep[Section 2.1.3]{overhaus2007equity}. The price process is not a martingale anymore if $\rho > 0$. In the context of the SLV model, \citet{ren2007calibrating} use a specific time-dependent $\theta$, function of $\kappa,\sigma$, such that the conditional expected variance is unity and the correlation $\rho$ is set to zero. Note that the direct volatility process reads $\diff v =  v \left[\kappa (\theta - \ln v ) - \sigma^2 / 2\right]\diff t + \sigma v \diff W_v$. 
	\item The lognormal volatility model of \citet{tataru2010stochastic} is very similar to the Scott model, except that the mean reversion is on $v$ instead of $\ln v$:
\begin{subequations}		\begin{align}
		\frac{\diff F}{F} &=  v  L(F,t)  \diff W_F\,,\\
		\diff v &= \kappa (\theta - v)\diff t + \sigma v \diff W_v\,,
	\end{align} \end{subequations}
	This corresponds to the mean-reverting SABR model with $\beta=1$ and to the regular SABR model when $\kappa = 0$. Again, martingality imposes $\rho \leq 0$.  In \cite{tataru2010stochastic}, the parameters $\kappa, \theta, \sigma, \rho$ are allowed to be piecewise-constant in time, although in practice, the choice $\theta=v(0)=1$ and $\kappa=1$ is made. The model is not affine, and short time expansions may be used to approximate the vanilla option prices for calibration. The stochastic volatility parameters are however not necessarily calibrated to vanilla options, as the volatility process is only supposed to capture the market behavior. While the volatility process may have some infinite moments, it is not an issue in practice, as the local volatility component will annihilate the explosion in practice.
\end{itemize}


\section{Classical implied parameters}
\subsection{Textbook approach}
The textbook approach to stochastic volatility calibration is to find the parameters which minimize the error in weighted vanilla option prices of various maturities and strikes. 
The options chosen are typically the ones used to build the volatility surface, that is, the ones used for hedging the trader's book. Those parameters are then used to price various exotics, typically by Monte-Carlo simulation. For the Heston model, the five parameters are calibrated to all the vanilla options as described in \citep[Section 2.1.1]{overhaus2007equity}, as well as in \citep[Section 4.7]{zhu2009applications} and also in \citep{mikhailov2004heston,lefloch2014fourier}. They thus represent the full volatility surface. In context of FX options \citet{griebsch2010stochastic} also calibrate the Heston model to the relevant set of options of various strikes and maturities to price a barrier option. For the barrier option with maturity 18 months (18M) considered, this means the cross product of the 1M, 2M,  3M, 6M, 9M, 1Y, 2Y tenors and 10-d put/call, 25-delta put/call, ATM vanilla market implied volatilities.
A single set of Heston parameters is produced as output.

\subsection{Fixing one parameter}
A small practical refinement  consists in not calibrating one of the parameters, typically the speed of mean reversion $\kappa$ to vanilla options, but have it set exogenously.  This has the advantage to stabilize the calibration \citep[Section 4.7.2]{zhu2009applications}, because the long term variance $\theta$ and the speed of mean reversion $\kappa$ play a similar role \citep{buehler2004stochastic}. \citet{bergomi2004smile} lets $\kappa=2$ because the dynamics of the variance is mostly reflected in that of short vols, corresponding to maturities $T$ such that $T \ll \frac{1}{\kappa}$. \citet{zhu2009applications} also chooses the relatively large value $\kappa = 2$, with the idea that the Feller condition $2\kappa\theta \ge \sigma^2$ is more likely to hold. In the context of his variance curve model,  \citet[Proposition 3.1]{buehler2006consistent} proves that the Heston model is not free of dynamic arbitrage if $\kappa$ is not kept constant.

Another important motivation to fix some of the parameters lies in the fact that vanilla option prices do not capture the forward volatility dynamic at all. And thus calibrating the model only on the vanilla option prices may lead to the wrong dynamic. Instead, one may fix the speed of mean reversion or the vol of vol based on exotic options or forward starting options prices.

It is not uncommon to set the initial volatility (or variance for Heston) $v_0$ from the implied volatility of the ATM vanilla option of maturity 1M  \citep[Section 4.7.3]{zhu2009applications}. This further speeds up and stabilizes the calibration.

\subsection{Further stabilization}
When the Heston model is recalibrated, for example every business day, the Heston parameters will change each time. Frequent large variations increase the risk of bad hedges and mean that the model is somewhat unstable. \citet{andersen2010interest1, buehler2004stochastic} propose to add a penalty to the minimization, such that the parameters do not deviate too much from their previously calibrated values. The challenge is to choose the correct penalty value. 

\citet{buehler2004stochastic} suggests to compute firstly the error of the unpenalized calibration and then to add a penalty such that the total error is twice the unpenalized error. This is not necessarily always appropriate, for example if the unpenalized calibration error is large then the penalty allowed is large and we force the parameters to stay close to their previous values. If the unpenalized error is small, the new parameters may move far from their previous values. Furthermore, in the case of a disruptive market event, the parameters must be able to move far.

Alternatively, one may perform a local minimization starting from the previously calibrated parameters (except possibly the initial variance). It is of course not guaranteed that the result does not end up too far in some cases, unless we impose box constraints on the parameters. Similarly, the local minimization may start from a geometrical initial guess, based purely on the implied volatilities at a three different strikes and two maturities, typically using a small time, or near-moneyness approximation \citep{forde2012small}.

\subsection{In the context of the SLV model}
In reality, the calibration towards the full implied volatility surface is rarely used in the context of an SLV model for several reasons:
 \begin{itemize}
 	\item The local volatility component already offers a perfect fit to the vanilla options.
 	\item The set of parameters is often reduced as a consequence of the first point and the stochastic volatility part is there to capture the dynamic, which is not necessarily visible in vanilla option prices.
 	\item The mixing weight is usually a maturity dependent term-structure. And thus, some of the effective stochastic volatility parameters are actually different for each maturity.
 \end{itemize}
It may however still makes sense if the stochastic volatility model uses time varying parameters in its volatility process, although this is explicitly not how the model of \cite{tataru2010stochastic} is calibrated.
The calibration towards the full implied volatility surface may also be used when the calibration view is global, i.e. not specific to certain products such as barrier options or cliquets \citep{qu2016manufacturing}.

\subsection{Variance swap calibration}
The variance swap term-structure offers another interesting way to calibrate some of the stochastic volatility parameters. For example, in the Heston model, the expected total variance is known in closed-form, and depends only on the speed of mean reversion $\kappa$, the mean reversion value $\theta$ and the initial variance $v_0$:
\begin{equation}
	\frac{1}{T}\mathbb{E}\left[v(T)\right] = \theta  + \frac{v_0-\theta}{\kappa T}\left(1-e^{-\kappa T}\right)\,.
\end{equation}
 We may thus calibrate those parameters either to the term structure of quoted variance swaps, or to the term-structure of (at least three) variance swap prices implied by the vanilla options, using the replication formula of \citet{carr2001towards}. \citet{guillaume2014heston} use those parameters as initial guess for a  calibration towards all vanilla options prices, and this may help to stabilize such a calibration. It may however make more sense, especially in the context of SLV models, to let those parameters fixed by the variance swap term-structure. The practice is applicable to most stochastic volatility models, and is reminiscent of the forward variance curve approach (see Section \ref{sec:variance_curve}).
 There is however a major issue that arises in practice: models like the classic Heston with constant parameters can not always fit the term-structure of variance swaps, as its shape is very constrained, and may thus lead to financially nonsensical parameters \citep{lefloch2025calibrating}. The two factors Heston model fares better in this regard. And the Heston model with piecewise-constant parameters and $\mathbb{E}[v(T)]=1$ is essentially a variance curve model, and may be perfectly calibrated through $L(T)$ (although it may be more judicious to calibrate it to vanillas directly).


\section{Maturity dependent parameters}\label{sec:maturity_dependent}
The basic idea behind maturity dependent parameters is to use the Heston model like the Black-Scholes model. In the Black-Scholes model, the implied volatility is a function of time and strike. Practitioners price exotic contracts such as barrier options, with a single volatility, chosen to be the at-the-money volatility at the time to maturity. This allows the use of the closed-form formulas (with many caveats obviously). In similar fashion, the parameters of the Heston model may be considered as a five dimensional implied volatility. The difference is that for a given maturity, the parameters are the same for any option strike. Thus the model is used with parameters that vary per maturity, but the model itself consists of constant parameters. This is not very consistent, akin to the varying implied volatility used in the Black-Scholes model. Maturity dependent parameters are common in the interest rate derivatives world, for example, this is how the market-standard SABR model of \citet{hagan2002managing} is used for swaption pricing. In general, the SABR model is however used this way only for smile interpolation, and rarely to price strongly path-dependent exotics.

In the Black-Scholes world, a slightly more elaborate practice is to consider the term-structure of at-the-money volatilities (often at the cost of being able to use analytical formulas). This would be akin to the use of a time-dependent Heston model, where, typically, the parameters are piecewise-constant and are calibrated in a bootstrap manner to all option maturities. The model itself takes into account internally the time-dependence in the stochastic volatility process evolution. This is not what we describe in this section.

In \citep[Tables 6.3 and 6.5]{clark2011foreign}, the Heston model is calibrated to vanilla FX options, option tenor by option tenor letting the initial variance $v_0$ be equal to the mean reversion level $\theta$ (called $m$ in the tables). We give an excerpt in Table \ref{tbl:clark_heston_calib}.
\begin{table}[h]
	\caption{FX volatilities and calibrated Heston parameters as of 16 September 2008, excerpt of \citet[Table 6.3]{clark2011foreign}. The Heston Feller ratio $\beta < 1$ shows that the Feller condition does not hold.  \label{tbl:clark_heston_calib}}
	\centering{
	\begin{tabular}{llrrrrrrrrr}\toprule
	ccypair & tenor & ATM & 25-d-MS & 25-d-RR & $v_0$ & $\rho$ & $\sigma$ & $\kappa$ & $\theta$ & $\beta$ \\ \midrule 
	EURUSD & 3M & 12.70\% & 0.28\% & -0.55\% & 0.02 & -0.13 & 0.49 & 6.02 & 0.02 & 0.90\\
	EURUSD & 6M & 11.87\% & 0.38\% & -0.55\% & 0.02 & -0.13 & 0.41 & 3.02 & 0.02 & 0.59\\
	EURUSD & 1Y & 11.50\% & 0.40\% & -0.55\% & 0.02 & -0.13 & 0.31 & 1.50 & 0.02 & 0.49\\
	EURUSD & 2Y & 11.45\% & 0.40\% & -0.55\% & 0.02 & -0.14 & 0.20 & 0.75 & 0.02 & 0.56\\				
	EURUSD & 3Y & 11.30\% & 0.40\% & -0.55\% & 0.02 & -0.15 & 0.16 & 0.50 & 0.02 & 0.55\\	
	EURUSD & 4Y & 11.13\% & 0.40\% & -0.56\% & 0.01 & -0.16 & 0.14 & 0.38 & 0.01 & 0.54\\	
	EURUSD & 5Y & 10.75\% & 0.38\% & -0.55\% & 0.01 & -0.17 & 0.12 & 0.30 & 0.01 & 0.56\\ \bottomrule
	\end{tabular}}
\end{table}
In particular, for each tenor, $\kappa$ is set using the rule of thumb $\kappa = 1.5 / T$, and the three remaining parameters $v_0, \rho, \sigma$ are calibrated to the at-the-money (ATM), 25 delta strangle (25-d-MS), 25 delta risk reversal (25-d-RR) market quotes as of September 16, 2008. \citet[Section 7.3]{clark2011foreign} provides different rules for the level and the speed of mean reversion: $\theta = \sigma_{\textsf{ATM}}^2(T)$ (the tables of  \cite[Section 6]{clark2011foreign} show $\theta = v_0$ instead), $\kappa = 2.75 / T$. In practice, the speed of mean reversion $\kappa$ will be an exogenous tenor-dependent parameter, adjusted by the trader or quant.

A similar approach is described in \citep{janek2011fx}: for a given vanilla option maturity, the speed of mean reversion $\kappa$ and the initial variance $v_0$ are fixed, and the three remaining parameters are calibrated to the set of implied volatilities corresponding to a set of delta pillars. Then exotic options, such as one-touch are priced with the parameters of the corresponding maturity. 
Somewhat interestingly, the authors precise that the prices obtained are remarkably close to the market prices in practice, while in \citep{wystup2020mixed}, Uwe Wystup asserts that the Heston model does not perform satisfactorily for any product. This is perhaps not as contradictory as it seems since the markets may have changed their practices significantly since 2011.

\citet[Section 7.7]{austing2014smile} gives the example of calibrating the SABR model using three vanilla options of same maturity as the exotic contract, including the ATM option, see also \citep{lefloch2014explicit}. The SABR $\alpha$ is set such that the ATM option price is exactly reproduced. 

In the context of the Tremor SLV model \citep{wystup2011tremor}, which consists of the Heston stochastic volatility model augmented with a quadratic local volatility function, the mean reversion parameters are again fixed such that $\kappa=1$ and $\theta=v_0$. For a given maturity, The three stochastic volatility parameters are calibrated to 10-delta put/call, and ATM options market quotes. The local component parameters are calibrated to all options at this maturity after the mixing weight has been applied to the vol of vol $\sigma$.

\section{Term-structure of parameters}
With an affine stochastic volatility model, it is still possible to price vanilla options with piecewise-constant parameters in time as shown in \citep{mikhailov2004heston,elices2009affine}. The stochastic volatility process takes into account the variation of those parameters in time. The resulting pricing formula is very similar, the additional complexity lies in the recursive form of the characteristic function, which is then also more costly to evaluate, with a cost linear in the number of terms in the structure.


For non affine models, the parameter averaging technique of \cite[Section 9.3]{andersen2010interest1} provides ways to approximate vanilla option prices of stochastic volatility models with a term-structure of parameters.

This obviously provides a better fit, without a full blown SLV model. The term-structure of parameters is appealing but we end up with a lot of parameters, which increases the chance of instabilities of the model in a day to day usage \citep{overhaus2007equity}: aren't we overfitting? how stable are those parameters from day to day? how would you hedge against the variation of all those parameters?

In reality, a term-structure of parameters may be more relevant to SLV models, as the mixing weight will introduce such a term-structure. 

\section{SLV practices and the mixing parameter}
In \cite{ren2007calibrating}, the two free stochastic volatility  parameters of the model, the vol of vol and the speed of mean reversion, are set exogenously (not calibrated to vanillas). The mean reversion is implicit in the model and the correlation is zero. The stochastic volatility is thus not calibrated to vanilla options at all, and is only there to add an interesting dynamic to the local volatility. This makes this SLV model simpler to understand and risk manage. There is no need for a mixing parameter as the two stochastic volatility parameters are in effect the mixing. The two parameters may be captured as a term-structure (although in the stochastic volatility process, they are constant), to allow for a fine grained control over exotic prices across various maturities or tenors. 

In the context of the equity derivatives market, \citet{qu2016manufacturing} recommends against setting the correlation to zero, even if the local volatility component already contains the spot dependence, because the spot and volatility correlation is economically apparent. The correlation would then be set from a historical analysis. In practice, this is not so obvious: the stochastic volatility is not observable and a proxy must be used to estimate the correlation. The proxy may be the VIX (the SPX500 volatility index) price or the short term realized volatility of the underlying asset. It is however only a proxy. Such an historical correlation does not directly correspond to the stochastic volatility model correlation and must be adjusted accordingly. Furthermore, if we use the SLV model to generate the SPX index paths, the local volatility part will also generate a correlation between a (virtual) VIX index and the SPX index.  

In \cite{tataru2010stochastic}, only the two stochastic volatility parameters $\sigma$ and $\rho$ are term-structures of piecewise-constant values (along the time to maturity). They are set together through a mixing factor $\lambda$, following the rule $\sigma = \lambda \sigma_{\max}$, $\rho = \lambda \rho_{\max}$ where $\sigma_{\max}, \rho_{\max}$ are constants (maximally stochastic parameters). The mixing factor is also piecewise-constant in time and is effectively the only time-varying parameter the user of this SLV model sets. The stochastic volatility parameters are calibrated infrequently to the market, not to vanilla options, but to the historical dynamic of the volatility skew and spot. As the stochastic volatility parameters are not calibrated to vanillas, it does not matter if, in the model, vanilla options are priced either with slow numerical methods or fast but not so accurate approximations.

For the Heston SLV model \citet{clark2011foreign} applies a mixing weight to mark down strangles and risk reversals of the maturity considered (the corresponding market quotes are multiplied by the mixing weight) to reduce the convexity of the pure stochastic volatility part. The stochastic volatility parameters are then directly calibrated to those marked-down quotes. This translates to a lower vol of variance $\sigma$ but also to a lower correlation $\rho$, and a slight change of the initial variance $v_0$. When the mixing weight is zero, we expect the correlation and the vol of variance to be zero as well, the model will behave like a pure local volatility model. The mixing weight is tuned such that the price of double no touch options are close to the market prices for a given tenor. 

For the Tremor SLV model, the mixing weight is used to adjust the vol of vol $\sigma$, after the pure Heston stochastic volatility parameters $v_0, \rho, \sigma$ have been calibrated to vanilla options.

\citet{austing2014smile} shows that the correlation $\rho$ has little impact on the price of barrier options, and may be set to zero as the local volatility component will compensate for it. Furthermore the price of barrier options is mostly dependent on the vol of vol parameter. As a consequence, in the lognormal SLV model, for a given maturity, only the vol of vol is tuned to match the market prices of barrier options, such as one-touch contracts, via a mixing parameter $\lambda$ and the effective vol of vol used in the full-blown SLV model is $(1-\lambda)\sigma$ \citep[Section 9.6]{austing2014smile}. The stochastic volatility parameters are calibrated (approximately) to vanilla options of a given maturity.


In general, the mixing weight can not be constant for all maturities in order to fit the quoted prices of various barrier options (such as touch, no-touch).

\section{Live calibration vs. up-front calibration}
When does the stochastic volatility model calibration occurs?
There are two practices.

The first one, perhaps more traditional is the up-front calibration where the model is calibrated once a day or less. The parameters are stored, and then used to price exotics of various maturities during the day. This may also be applied more generally to SLV models.

The second practice, which may also motivate the choice of maturity dependent parameters, is to calibrate (some of) the stochastic volatility parameters at each valuation time. The parameters do not need to be stored and will be different at a later valuation time. When maturity-dependent parameters are used, only one maturity is calibrated, the one relevant for the exotic contract. In the context of the Heston model, the usual practice is to only calibrate the three parameters $v_0, \rho, \sigma$ (see Section \ref{sec:maturity_dependent}). In the context of a first generation SLV model with a parametric LV function, the LV parameters are also calibrated at the same time (but typically just after the SV parameters have been calibrated). This approach allows for a fast calibration.

It is perfectly possible to perform a live calibration towards a set of option contracts (not necessarily of the same maturity) although, those would typically be vanilla option contracts. The choice of vanilla (hedging) options may be different for different exotic trades. This makes a priori more sense: for example a two year exotic contract paying every 6 months will not be hedged only with the two year vanilla options.

In the end, while the up-front approach may seem more consistent during the day, it will lose this consistency the next day, when the parameters are recalibrated. When the parameters are calibrated towards the full set of vanilla options (classical implied calibration), there is however no good reason to perform a live calibration as long as the set of reference  vanilla options quotes has not moved. When embedded in an SLV model, the SV parameters may be kept constant for a longer period of time.





\funding{This research received no external funding.}
\conflictsofinterest{The authors declare no conflict of interest.}
\externalbibliography{yes}
\bibliography{choosing_stochastic_vol_parameters.bib}

\begin{thebibliography}{}

\bibitem[\protect\citeauthoryear{Andersen and Piterbarg}{Andersen and Piterbarg}{2010}]{andersen2010interest1}
Andersen, Leif and Vladimir Piterbarg. 2010.
\newblock Interest rate modeling--volume i.
\newblock {\em Atlantic Financial Press\/}~{\em 1}.

\bibitem[\protect\citeauthoryear{Austing}{Austing}{2014}]{austing2014smile}
Austing, Peter. 2014.
\newblock {\em Smile pricing explained}.
\newblock Springer.

\bibitem[\protect\citeauthoryear{Bergomi}{Bergomi}{2004}]{bergomi2004smile}
Bergomi, Lorenzo. 2004.
\newblock Smile dynamics i.
\newblock {\em Available at SSRN 1493294\/}.

\bibitem[\protect\citeauthoryear{Buehler}{Buehler}{2004}]{buehler2004stochastic}
Buehler, Hans. 2004.
\newblock Stochastic volatility models and products.
\newblock {\em Presentation at the Risk Training Course, Hong Kong\/}.

\bibitem[\protect\citeauthoryear{Buehler}{Buehler}{2006}]{buehler2006consistent}
Buehler, Hans. 2006.
\newblock Consistent variance curve models.
\newblock {\em Finance and Stochastics\/}~{\em 10\/}(2), 178--203.

\bibitem[\protect\citeauthoryear{Carr and Madan}{Carr and Madan}{2001}]{carr2001towards}
Carr, Peter and Dilip Madan. 2001.
\newblock Towards a theory of volatility trading.
\newblock {\em Option Pricing, Interest Rates and Risk Management, Handbooks in Mathematical Finance\/}, 458--476.

\bibitem[\protect\citeauthoryear{Christoffersen, Heston, and Jacobs}{Christoffersen et~al.}{2009}]{christoffersen2009shape}
Christoffersen, Peter, Steven Heston, and Kris Jacobs. 2009.
\newblock The shape and term structure of the index option smirk: Why multifactor stochastic volatility models work so well.
\newblock {\em Management Science\/}~{\em 55\/}(12), 1914--1932.

\bibitem[\protect\citeauthoryear{Clark}{Clark}{2011}]{clark2011foreign}
Clark, Iain~J. 2011.
\newblock {\em Foreign exchange option pricing: a practitioner's guide}.
\newblock John Wiley \& Sons.

\bibitem[\protect\citeauthoryear{Elices}{Elices}{2009}]{elices2009affine}
Elices, Alberto. 2009.
\newblock Affine concatenation.
\newblock {\em Wilmott Journal: The International Journal of Innovative Quantitative Finance Research\/}~{\em 1\/}(3), 155--162.

\bibitem[\protect\citeauthoryear{Forde, Jacquier, and Lee}{Forde et~al.}{2012}]{forde2012small}
Forde, Martin, Antoine Jacquier, and Roger Lee. 2012.
\newblock The small-time smile and term structure of implied volatility under the heston model.
\newblock {\em SIAM Journal on Financial Mathematics\/}~{\em 3\/}(1), 690--708.

\bibitem[\protect\citeauthoryear{Griebsch and Pilz}{Griebsch and Pilz}{2010}]{griebsch2010stochastic}
Griebsch, Susanne~A and Kay~F Pilz. 2010.
\newblock Stochastic volatility models: Foreign exchange.
\newblock {\em Encyclopedia of Quantitative Finance\/}.

\bibitem[\protect\citeauthoryear{Guillaume and Schoutens}{Guillaume and Schoutens}{2014}]{guillaume2014heston}
Guillaume, Florence and Wim Schoutens. 2014.
\newblock Heston model: the variance swap calibration.
\newblock {\em Journal of Optimization Theory and Applications\/}~{\em 161\/}(1), 76--89.

\bibitem[\protect\citeauthoryear{Guyon and Henry-Labordere}{Guyon and Henry-Labordere}{2011}]{guyon2011smile}
Guyon, Julien and Pierre Henry-Labordere. 2011.
\newblock The smile calibration problem solved.
\newblock {\em Available at SSRN 1885032\/}.

\bibitem[\protect\citeauthoryear{Hagan, Kumar, Lesniewski, and Woodward}{Hagan et~al.}{2002}]{hagan2002managing}
Hagan, Patrick~S, Deep Kumar, Andrew~S Lesniewski, and Diana~E Woodward. 2002.
\newblock Managing smile risk.
\newblock {\em Wilmott magazine\/}.

\bibitem[\protect\citeauthoryear{Hagan, Lesniewski, and Woodward}{Hagan et~al.}{2018}]{hagan2018implied}
Hagan, Patrick~S, Andrew~S Lesniewski, and Diana~E Woodward. 2018.
\newblock Implied volatility formulas for heston models.
\newblock {\em Wilmott\/}~{\em 2018\/}(98), 44--57.

\bibitem[\protect\citeauthoryear{Healy}{Healy}{2025}]{healy2025applied}
Healy, Jherek. 2025.
\newblock {\em Applied Quantitative Finance for Equity Derivatives\/} (Fifth ed.).
\newblock available Amazon.com and other online stores.
\newblock ISBN: {979-8289924087}.

\bibitem[\protect\citeauthoryear{Heston}{Heston}{1993}]{heston1993closed}
Heston, Steven~L. 1993.
\newblock A closed-form solution for options with stochastic volatility with applications to bond and currency options.
\newblock {\em Review of financial studies\/}~{\em 6\/}(2), 327--343.

\bibitem[\protect\citeauthoryear{Janek, Kluge, Weron, and Wystup}{Janek et~al.}{2011}]{janek2011fx}
Janek, Agnieszka, Tino Kluge, Rafa{\l} Weron, and Uwe Wystup. 2011.
\newblock Fx smile in the heston model.
\newblock In {\em Statistical tools for finance and insurance}, pp.\  133--162. Springer.

\bibitem[\protect\citeauthoryear{{Le Floc'h}}{{Le Floc'h}}{2014}]{lefloch2014fourier}
{Le Floc'h}, Fabien. 2014.
\newblock Fourier integration and stochastic volatility calibration.
\newblock {\em Available at SSRN 2362968 \url{https://ssrn.com/abstract=2362968}\/}.

\bibitem[\protect\citeauthoryear{{Le Floc'h}}{{Le Floc'h}}{2025}]{lefloch2025calibrating}
{Le Floc'h}, Fabien. 2025.
\newblock Calibrating {Heston} to variance swaps - a bad idea?
\newblock \url{https://chasethedevil.github.io/post/heston_variance_swap_calibration/} [Online; accessed 12-February-2025].

\bibitem[\protect\citeauthoryear{{Le Floc'h} and Kennedy}{{Le Floc'h} and Kennedy}{2014}]{lefloch2014explicit}
{Le Floc'h}, Fabien and Gary~J Kennedy. 2014.
\newblock Explicit {SABR} calibration through simple expansions.
\newblock {\em Available at SSRN 2467231\/}.

\bibitem[\protect\citeauthoryear{Mikhailov and N{\"o}gel}{Mikhailov and N{\"o}gel}{2004}]{mikhailov2004heston}
Mikhailov, Sergei and Ulrich N{\"o}gel. 2004.
\newblock {\em Heston’s stochastic volatility model: Implementation, calibration and some extensions}.
\newblock John Wiley and Sons.

\bibitem[\protect\citeauthoryear{Overhaus, Berm{\'u}dez, Buehler, Ferraris, Jordinson, and Lamnouar}{Overhaus et~al.}{2007}]{overhaus2007equity}
Overhaus, Marcus, Ana Berm{\'u}dez, Hans Buehler, Andrew Ferraris, Christopher Jordinson, and Aziz Lamnouar. 2007.
\newblock {\em Equity hybrid derivatives}, Volume 374.
\newblock John Wiley \& Sons.

\bibitem[\protect\citeauthoryear{Qu}{Qu}{2016}]{qu2016manufacturing}
Qu, Dong. 2016.
\newblock Manufacturing and managing customer-driven derivatives.
\newblock {\em The Wiley finance series\/}.

\bibitem[\protect\citeauthoryear{Ren, Madan, and Qian}{Ren et~al.}{2007}]{ren2007calibrating}
Ren, Yong, Dilip Madan, and M~Qian Qian. 2007.
\newblock Calibrating and pricing with embedded local volatility models.
\newblock {\em RISK-LONDON-RISK MAGAZINE LIMITED-\/}~{\em 20\/}(9), 138.

\bibitem[\protect\citeauthoryear{Sch{\"o}bel and Zhu}{Sch{\"o}bel and Zhu}{1999}]{schobel1999stochastic}
Sch{\"o}bel, Rainer and Jianwei Zhu. 1999.
\newblock Stochastic volatility with an {Ornstein--Uhlenbeck} process: an extension.
\newblock {\em European Finance Review\/}~{\em 3\/}(1), 23--46.

\bibitem[\protect\citeauthoryear{Scott}{Scott}{1987}]{scott1987option}
Scott, Louis~O. 1987.
\newblock Option pricing when the variance changes randomly: Theory, estimation, and an application.
\newblock {\em Journal of Financial and Quantitative analysis\/}~{\em 22\/}(4), 419--438.

\bibitem[\protect\citeauthoryear{Tataru and Fisher}{Tataru and Fisher}{2010}]{tataru2010stochastic}
Tataru, Grigore and Travis Fisher. 2010.
\newblock Stochastic local volatility.
\newblock {\em Quantitative Development Group, Bloomberg Version\/}~{\em 1\/}(February 5).

\bibitem[\protect\citeauthoryear{Wystup}{Wystup}{2011}]{wystup2011tremor}
Wystup, Uwe. 2011.
\newblock The tremor stochastic-local-volatility model: Independent validation by math-finance.
\newblock In {\em 2011 Global Derivatives USA Conference}.

\bibitem[\protect\citeauthoryear{Wystup}{Wystup}{2020}]{wystup2020mixed}
Wystup, Uwe. 2020.
\newblock Mixed local volatility boosts distribution of exotics.
\newblock {\em Wilmott\/}~{\em 2020\/}(110), 34--37.

\bibitem[\protect\citeauthoryear{Zhu}{Zhu}{2009}]{zhu2009applications}
Zhu, Jianwei. 2009.
\newblock {\em Applications of {Fourier} transform to smile modeling: Theory and implementation}.
\newblock Springer Science \& Business Media.

\end{thebibliography}
\appendixtitles{no}
\end{document}